# Reflexões sobre o negacionismo da curvatura da Terra a partir da participação do público na reprodução coletiva do experimento de Eratóstenes

# Reflections on the denialism of the Earth's curvature based on the general public participation in the collective reproduction of Eratosthenes' experiment

Karen Luz Burgoa Rosso[1], José Alberto Casto Nogales Vera[2], Ana Eliza Ferreira Alvim Silva[3]

**Resumo**

Neste trabalho, propomos a replicação de experimentos científicos, com a participação do público não especializado, como uma das estratégias possíveis para o enfrentamento do negacionismo científico. Pelas redes sociais da Universidade Federal de Lavras (UFLA), o projeto de extensão "A Magia da Física e do Universo" convidou a população a reproduzir o experimento feito pelo grego Eratóstenes há mais de dois mil e cem anos, por meio do qual ele comprovou que a Terra é esférica. Apresentamos o relato dessa experiência para sugerir que, de forma simples, é possível ao cidadão experimentar o conhecimento científico e observar os fenômenos para uma interpretação crítica das informações falsas que circulam na sociedade. Essa proposta tem o diferencial de buscar desconstruir o negacionismo da curvatura da Terra por meio da participação direta e coletiva dos cidadãos no experimento de Eratóstenes, o que permite a reflexão sobre como funciona o método científico envolvido no estudo dessa temática.

**Palavras-chave:** experimento de Eratóstenes; curvatura da Terra; negacionismo da ciência.

[1] Professora da Universidade Federal de Lavras (UFLA), Departamento de Física/Instituto de Ciências Naturais (DFI/ICN) e do posgrado em Educação científica e ambienta PPGECA/UFLA. http://lattes.cnpq.br/5146304381504697. E-mail: karenluz@ufla.br

[2] Professor da Universidade Federal de Lavras (UFLA), Departamento de Física/Instituto de Ciências Naturais (DFI/ICN)e do posgrado em Educação científica e ambienta PPGECA/ UFLA. . http://lattes.cnpq.br/7972689654356212. E-mail: jnogales@ufla.br

[3] Coordenadora de Divulgação Científica da Universidade Federal de Lavras (UFLA) e jornalista da Coordenadoria de Comunicação Social UFLA. http://lattes.cnpq.br/8530067620833656. E-mail: anaeliza.alvim@ufla.br

**Abstract**

In this paper, we propose the replication of scientific experiments, with the participation of the general public, as one of the possible strategies to confront science deniers. Using the social media of the Federal University of Lavras (UFLA), the project entitled "The Magic of Physics and the Universe" invited the public to reproduce the experiment carried out by the Greek Eratosthenes over 2,100 years ago, which he proved that the Earth is spherical. We present a report of this experience to suggest that it is possible for citizens to experience scientific knowledge and observe the phenomena in a simple way, in order to develop a critical interpretation of false information that circulates in society. This proposal distinctively geeks to deconstruct the denial of the Earth's curvature through direct and collective participation of citizens in Eratosthenes' experiment, which permit more reflection on how the scientific method used to study this subject works.

**Introdução**

Mesmo em tempos considerados sombrios, como a Idade Média, sonhadores continuaram explorando o cosmos, aspirando a uma descoberta individual e coletiva, intelectual e livre do nosso universo, contrapondo-se a todo movimento obscurantista. Estamos vivendo um desses tempos, um momento distópico global, que nasce da confluência de muitas crises. À crise de saúde pública, provocada pela pandemia de Covid-19, somam-se a crise econômica decorrente de décadas e uma crise cultural global.

Em meio a essas crises atuais, colisões entre buracos negros confirmaram a realidade do espaço-tempo concebido por Albert Einstein, comemorou-se cem anos da confirmação da Teoria da Relatividade Geral de Albert Einstein e a primeira foto de um buraco negro foi apresentada.

Apesar de conquistas evidentes da ciência, na atualidade, há uma ofensiva anticientífica e obscurantista. Movimentos irracionalistas, terraplanistas, anti vacinas, revisionistas históricos, racistas, misóginos, entre outros, têm em comum o combate à razão, à ciência e às diversas formas de negação da realidade.

Assim como não podemos enxergar, no dia a dia, a olho nu, o vírus causador da Covid-19, que parou a humanidade e causou tantas mortes, também não conseguimos, com nossos sentidos, visualizar diretamente, no cotidiano, a esfericidade da Terra, sentir seu movimento, ou perceber o tamanho real do sol. Os cientistas nos informam sobre tudo isso, mas a experiência individual cotidiana direta pode não nos confirmar essas informações. Este é, por exemplo, um dos recursos que os movimentos como os dos terraplanistas utilizam, refutando as informações científicas sobre a esfericidade do planeta a partir do que nossos sentidos conseguem ou não conseguem captar [7].

A ciência, no entanto, utiliza métodos e instrumentos que permitem observações e conclusões que estão além da nossa percepção direta do dia a dia. É com a ajuda de microscópios que se pode visualizar microrganismos, e com aceleradores se constata partículas como os quarks e a partícula de Higgs, recentemente descoberta. O fato de não os vermos a olho nu não é evidência de que não existam. Da mesma forma, o fato de não irmos ao espaço observar o formato arredondado da Terra não significa que ela não possa sê-lo. A ciência trabalha com o método científico para a construção do conhecimento e apresenta ao senso comum um raciocínio lógico; a análise das experiências busca ser reflexiva, analítica e crítica, para analisar a realidade natural ou social. Enquanto o senso comum geralmente se baseia nos dados imediatos, à procura de explicações, o conhecimento científico procura bases sólidas, justificações claras e exatas mais detalhadas.

Os cientistas não buscam apenas evidências que confirmem suas hipóteses; eles também buscam evidências que possam refutá-las, e tudo que produzem é avaliado e questionado por outros especialistas, num processo de escrutínio crítico. É legítimo que cidadãos também não devam aceitar com passividade as informações que recebem. Precisam ser críticos frente às informações científicas e às tecnologias desenvolvidas, mas devem também ser coerentes diante da ciência, que tantos avanços já trouxe à humanidade, como a energia elétrica e a fibra ótica, a erradicação da varíola, a edição do genoma, a construção de telescópios capazes de ver galáxias a trilhões de quilômetros de distância, entre tantos outros avanços.

Um caminho para que a popularização da ciência auxilie no combate ao negacionismo científico é o de utilizarmos estratégias que possam aprofundar as percepções do senso comum por meio da prática do método científico, de forma que o público não especializado possa acompanhar e compreender o processo de produção de conhecimento por meio da ciência, envolvendo-se diretamente em atividades relacionadas ao método científico, alcançando a percepção de que nem sempre a relação que temos com a realidade no cotidiano será a única forma de conhecer e compreender essa realidade. Embora também a observação cotidiana pelos sentidos colabore e faça parte da construção do conhecimento científico, é preciso conjugar outros recursos e instrumentos a essas observações na produção da ciência.

Nesse sentido, este texto relata uma iniciativa de popularização da ciência que propôs ao público das mídias sociais da Universidade Federal de Lavras (UFLA), localizada no sul de Minas Gerais, que participasse de uma reprodução coletiva do experimento feito pelo grego Eratóstenes há cerca de 2.100 anos atrás, por meio do qual ele conseguiu concluir que a Terra não poderia ser plana, e encontrou a medida da circunferência aproximada do planeta. Além da observação do tamanho de sombras projetadas pelo sol, ele precisou utilizar cálculos matemáticos e geométricos, além de outras reflexões científicas. Mesmo com recursos hoje bastante simples, ele chegou a um resultado muito próximo do que foi confirmado posteriormente por outras investigações científicas.

O objetivo de envolver o público na reprodução do experimento foi o de permitir às pessoas a compreensão sobre a forma como a ciência observa a realidade e interage com ela, exigindo procedimentos metódicos e fundamentados.

Monitoramos as respostas do público e apresentamos aqui as reflexões extraídas dessa experiência. Também constituímos materiais didáticos e ferramentas de cálculo que podem ser utilizados como parâmetro por outros grupos, inclusive para atividades na educação básica, para realização coletiva do experimento, de forma a estimular que o público não especializado reflita sobre posições negacionistas, como a que diz que a Terra seria plana.

A iniciativa foi conduzida pelo projeto de extensão da UFLA "A Magia da Física e do Universo", em parceria com as coordenadorias de Comunicação e de Divulgação Científica da UFLA. Experiências parecidas foram desenvolvidas em outros países e cidades brasileiras, conforme referências [2-3], [5-6], [8-12].

**Métodos**

Para uma explicação simplificada do experimento e das observações de Eratóstenes, podemos dizer que ele considerou ser o sol de um tamanho muito superior ao da Terra e, portanto, com raios paralelos ao incidirem sobre o planeta. Se a Terra fosse plana, os raios do sol incidindo paralelamente sobre lugares diferentes do planeta não projetariam sombra alguma nesses lugares, ou projetariam sombras de mesmo comprimento. Mas Eratóstenes obteve a informação de que em Siena, no primeiro dia do solstício de verão (21 de junho), um objeto perpendicular ao solo não projetava sombra, e observou que no mesmo instante, em Alexandria, a 800 quilômetros ao norte, havia sombra bem definida. Então, imaginando que a única explicação para o fenômeno seria um planeta esférico, ele utilizou cálculos matemáticos para definir o tamanho da circunferência da Terra.

Para os cálculos, ele conhecia um princípio matemático simples (Figura 1): de que duas linhas paralelas cortadas por uma linha inclinada produzem ângulos internos alternados que são de mesma medida. Considerado os raios do sol como paralelos e o objeto colocado no lugar em que há sombra como a linha inclinada que corta as linhas paralelas, ele usou o princípio matemático para descobrir o ângulo entre o raio de sol e o objeto, que era, portanto, o mesmo ângulo entre as duas cidades tendo o centro da Terra como vértice. A partir daí, buscando a medida da distância entre as duas cidades, conseguiu calcular a medida do raio entre as duas cidades e também a circunferência da Terra.

Esse também foi o princípio utilizado para a coleta de dados no experimento e realização de cálculos.

Figura 1 - Ilustração do princípio matemático utilizado nos cálculos.

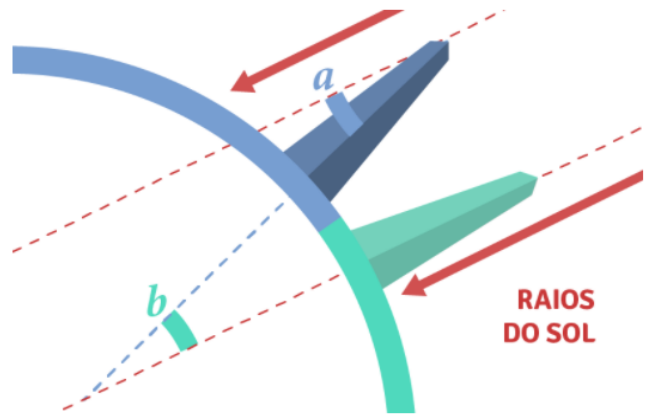

**Fonte:** GCFGlobal [4].

Para reprodução coletiva do experimento de Eratóstenes, divulgamos pelas diferentes mídias utilizadas pela UFLA as chamadas para que todos os interessados pudessem participar.

Figuras 2 a 4 - Exemplos de posts realizados para convidar o público para a participação no experimento coletivo.

Fonte: elaboração pela equipe do projeto com orientação dos autores.

Como a realização do experimento exige orientações mais detalhadas, para que tudo possa ser feito do modo correto, os interessados foram convidados a ingressar em um grupo de aplicativo de mensagens, pelo qual receberiam as atualizações sobre o evento e as orientações. O grupo chegou a reunir cerca de 130 pessoas, entre brasileiros e bolivianos.

Dois dias antes das datas programadas para realização conjunta do experimento, os professores responsáveis pela condução da iniciativa ministraram uma aula remota para discutir as reflexões científicas de Eratóstenes sobre a curvatura da Terra e sobre os passos que deveriam ser seguidos para a realização do experimento. Trinta e seis pessoas participaram desse encontro, que ficou gravado e foi disponibilizado para quem não acompanhou em tempo real. Mais de 95 visualizações do vídeo foram registradas posteriormente à aula, até o fechamento deste texto.

Durante a aula, os professores relataram a história sobre como Eratóstenes observou as sombras provocadas pelo sol e as reflexões que fez para constatar a curvatura da Terra. Foi utilizado o Stellarium (um software livre de astronomia para visualização do céu nos moldes de um planetário) para permitir que todos pudessem comparar a posição do sol em diferentes lugares do mundo e épocas do ano, inclusive simulando a situação do céu na data em que Eratóstene fez seu experimento. Foram também detalhados os procedimentos a serem feitos para a reprodução do experimento coletivamente. Os participantes apresentaram dúvidas e comentários. Assista ao vídeo da aula. O resumo das orientações para o experimento também foi disponibilizado no grupo formado no app de mensagens.

O ponto de referência em relação ao qual todos os participantes fizeram a medida foi local em que o sol ficou no zênite, com sombra zero ao meio-dia. Os autores fizeram suas medições no Centro de Convivência da UFLA, como forma simbólica de representar o experimento original, em que Alexandria foi uma referência. A Universidade representou, de modo análogo, um centro de circulação de conhecimento, como era a Biblioteca de Alexandria.

Os autores encontraram as coordenadas de latitude e longitude do referencial zero por meio do software Stellarium, e com Google Maps localizaram esse ponto, como mostram as elipses amarelas pontilhadas na figura 5. Na mesma figura, no lado direito pode ser observada a distância entre a referência zero e

o Centro de Convivência da UFLA, em frente à Biblioteca Universitária da UFLA.

Figura 5 - Passos para achar o ponto de referência zero e a distância até o Centro de Convivência da UFLA.

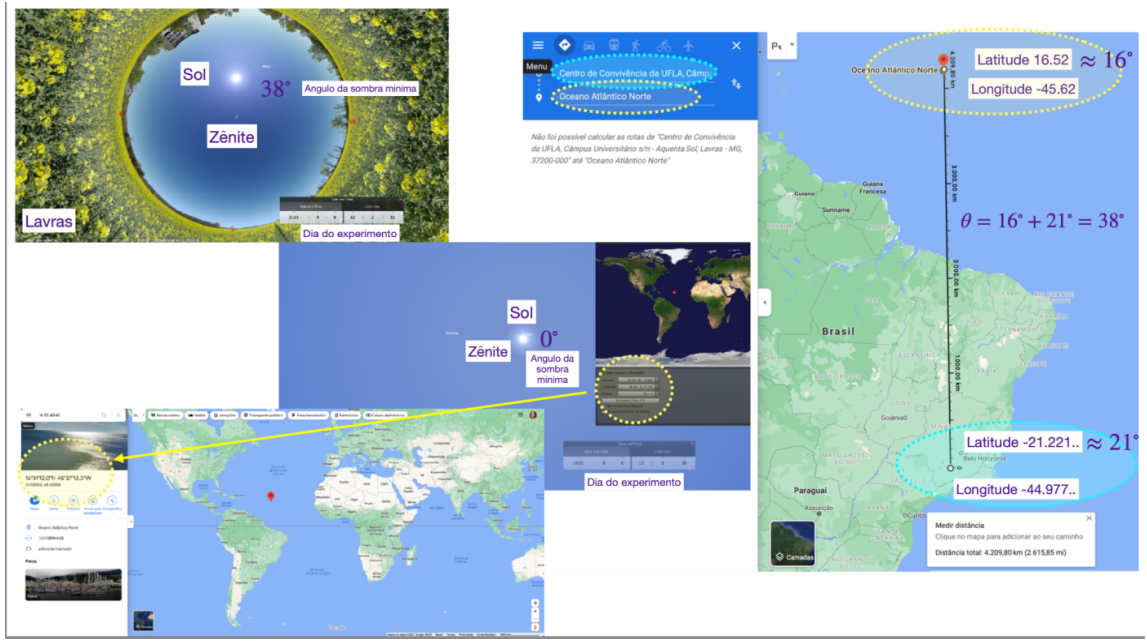

Fonte: dados coletados e processados pelos autores.

Nas datas programadas para a reprodução do experimento (6 e 7 de agosto de 2021), muitos participantes relataram céu nublado e chuva em seus locais de residência. Então foi necessário estender o número de dias programados para as medições (até 10 de agosto), de forma que todos tivessem a oportunidade de fazer as observações e registros.

Após o envio dos dados pelos participantes, por meio de formulário eletrônico preparado para tal, as informações foram processadas pela equipe do projeto, para constituição dos resultados e posterior divulgação aos participantes. O formulário foi elaborado de maneira a colher os dados técnicos e as impressões da vivência experimental, muito importantes para o entendimento dos significados que os números apresentam.

**Resultados**

Mesmo sendo um grupo formado eminentemente por professores e estudantes, 82,3% dos participantes que concluíram o experimento declararam que não haviam realizado antes essa observação prática, que é muito diferente de apenas ter acesso às informações teóricas. Esse dado nos mostra que a realização desse tipo de experimento pode introduzir novas experiências ao processo de acesso ao conhecimento.

No formulário, as pessoas enviaram os registros dos comprimentos das sombras de objetos que utilizaram para o experimento. Elas escolheram quantas vezes medir. A orientação foi a de começarem o experimento antes de o sol ficar a pino, anotando horário e valor de cada medição. Pedimos também a fotografia do experimento para relacionar a observação do ângulo da sombra com a distância entre a referência zero e o local onde as pessoas estavam realizando o experimento. Veja a figura 6, com a montagem dos dados enviados e seu significado na curvatura da Terra.

Figura 6 - Fotografias da investigadora Caroline, de Barbacena-MG, a primeira a enviar seus resultados. A coleta da sombra foi feita com um copo, resultando em um ângulo de 44 graus.

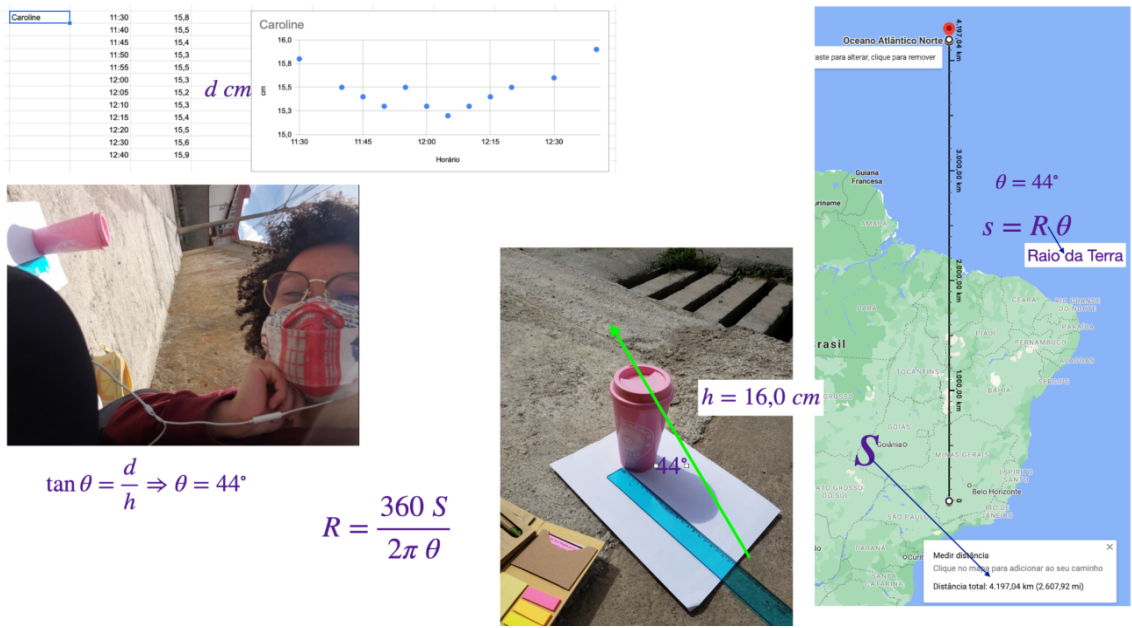

Fonte: arquivos coletados pelos autores durante o projeto.

Os participantes registraram a mudança no comprimento da sombra, e os dados registrados formaram uma forma curva que chamou a atenção da maioria deles, segundo relatos no formulário, ficando perceptível que a sombra vai reduzindo quando o sol está mais próximo da posição a pino, voltando a aumentar após esse período.

Os participantes informaram também a latitude e longitude do local do experimento. A diferença entre o valor absoluto da latitude do referencial zero e a latitude do local em que estava Caroline foi de 38 graus. Veja o valor na 12ª coluna da Figura 7, resultado bem próximo do ângulo de 44 graus, identificado com a sombra do copo que ela utilizou (7ª coluna). A medida do raio da Terra está na coluna 14: a Caroline obteve o valor aproximado de 5.524 km, com um erro percentual de 13,4%.

Figura 7 - Planilha com todos os dados e cálculos mais relevantes dos 23 participantes do experimento coletivo. Clique aqui para acessar a planilha e visualizar as fórmulas.

| Local | Nome | Altura objeto | Sombra mínima | d/h | erro d/h | Angulo | erro angulo | Distancia aproximada ao centro de convivencia da UFLA km | Latitude | Longitude | Valor absoluto de (Latitude - Referença) | Distancia aproximada do local ao ponto de referencia zero km | Resultado do raio da Terra km | erro devido a distancia (1km) km | erro devido ao angulo km | Erro percentual em relacao ao valor teorico do raio da terra |
|---|---|---|---|---|---|---|---|---|---|---|---|---|---|---|---|---|
| Ciudad de La Paz-provincia Murillo | Adrian | 5,5 | 3,0 | 0,55 | 0,03 | 29 | 1 | 2.492 | -16,498996 | -68,155582 | 33 | 3.673 | 7356 | 2 | 0,9 | 15,3% |
| La Paz, Bolivia | Gino | 22,0 | 12,2 | 0,555 | 0,007 | 29 | 0,3 | 2.497 | -16,480000 | -68,150000 | 33 | 3.666 | 7240 | 2 | 0,2 | 13,5% |
| Murillo, La Paz | Andrea | 18,0 | 16,0 | 0,89 | 0,01 | 42 | 0,3 | 2.491 | -16,4955455 | -68,1336229 | 33 | 3.677 | 5060 | 1 | 0,1 | 20,7% |
| La Paz- Bolivia | Clara | 19,0 | 17,0 | 0,89 | 0,01 | 42 | 0,3 | 3.130 | -16,4955455 | -68,1336229 | 33 | 3.682 | 5045 | 1 | 0,1 | 20,9% |
| Departamento de La Paz, provincia Murillo | Micaela | 7,1 | 4,5 | 0,63 | 0,02 | 32 | 0,9 | 2.490 | -16,487456 | -68,124192 | 33 | 3.677 | 6509 | 2 | 0,5 | 2,1% |
| La Paz, Bolivia | Israel | 79,5 | 49,0 | 0,616 | 0,002 | 32 | 0,1 | 2.487 | -16,496965 | -68,105411 | 33 | 3.673 | 6650 | 2 | 0,0 | 4,3% |
| La Paz, Bolivia | MIRKO | 90,2 | 54,8 | 0,608 | 0,002 | 31 | 0,1 | 2.496 | -16,550442 | -68,087326 | 33 | 3.678 | 6737 | 2 | 0,0 | 5,6% |
| La Paz - Bolivia | Dario | 31,1 | 20,3 | 0,653 | 0,005 | 33 | 0,2 | 2.482 | -16,544465 | -68,074333 | 33 | 3.673 | 6351 | 2 | 0,1 | 0,4% |
| Carrasco-Cochabamba-Bolivia | Tomas | 13,3 | 8,7 | 0,66 | 0,01 | 33 | 0,5 | 2.178 | -17,6741539 | -65,4194823 | 34 | 3.813 | 6563 | 2 | 0,3 | 2,9% |
| Santa Cruz/ Santa Cruz de la Sierra | Camila | 83,0 | 57,9 | 0,698 | 0,002 | 35 | 0,1 | 2.238 | -17,8334691 | -63,1763265 | 34 | 3.819 | 6270 | 2 | 0,0 | 1,7% |
| Marabá/Pará | Gabriela | 16,6 | 4,2 | 0,253 | 0,008 | 14 | 0,4 | 2.227 | -3,698600 | -49,462228 | 20 | 2.257 | 9108 | 4 | 0,7 | 42,8% |
| Mogi Mirim - São Paulo | Renata | 16,70 | 13,50 | 0,808 | 0,005 | 39 | 0,4 | 242 | -22,4014468 | -46,9523697 | 39 | 4.327 | 6365 | 1 | 0,2 | 0,2% |
| Guarulhos SP | James | 23,3 | 17,5 | 0,751 | 0,008 | 37 | 0,3 | 294 | -23,452879 | -46,517852 | 40 | 4.454 | 6914 | 2 | 0,1 | 8,4% |
| Caçapava / São Paulo | Murilo | 22,0 | 18,0 | 0,818 | 0,008 | 39 | 0,3 | 220 | -23,0927000 | -45,6975000 | 40 | 4.402 | 6419 | 1 | 0,1 | 0,6% |
| Varginha -MG | Leandro | 30,0 | 22,0 | 0,733 | 0,006 | 36 | 0,2 | 61 | -21,582849 | -45,422964 | 38 | 4.234 | 6691 | 2 | 0,1 | 4,9% |
| Lavras - MG | Danielle | 36,0 | 21,5 | 0,597 | 0,004 | 31 | 0,2 | 4 | -21,2475974 | -44,9968210 | 38 | 4.199 | 7799 | 2 | 0,1 | 22,3% |
| Lavras, Minas Gerais | Karina | 21,0 | 18,0 | 0,857 | 0,009 | 41 | 0,3 | 4 | -21,2224470 | -44,9919172 | 38 | 4.193 | 5917 | 1 | 0,1 | 7,2% |
| Lavras/MG | Ana Eliza | 9,6 | 7,8 | 0,81 | 0,02 | 39 | 0,7 | 1 | -21,2221098 | -44,9793107 | 38 | 4.205 | 6163 | 1 | 0,3 | 3,4% |
| Macuco de Minas - Minas Gerais | Valdeir | 27,8 | 22,3 | 0,802 | 0,006 | 39 | 0,2 | 33 | -21,278525 | -44,756598 | 38 | 4.205 | 6220 | 1 | 0,1 | 2,5% |
| Barroso - Minas Gerais | Steffani | 20,5 | 19,9 | 0,97 | 0,01 | 44 | 0,3 | 130 | -21,203063 | -43,970193 | 38 | 4.197 | 5447 | 1 | 0,1 | 14,6% |
| Barbacena,Minas Gerais | Caroline | 16,0 | 15,2 | 0,95 | 0,01 | 44 | 0,4 | 152 | -21,218300 | -43,774500 | 38 | 4.197 | 5524 | 1 | 0,1 | 13,4% |
| Barbacena- MG | Vitória | 11,5 | 8,5 | 0,74 | 0,02 | 36 | 0,6 | 152 | -21,216900 | -43,763900 | 38 | 4.190 | 6583 | 2 | 0,3 | 3,2% |
| Rio de Janeiro, Rio de Janeiro | BRUNO | 11,9 | 9,8 | 0,82 | 0,02 | 39 | 0,5 | 233 | -22,965522 | -43,708243 | 39 | 4.400 | 6387 | 1 | 0,2 | 0,1% |
| Centro de convivencia UFLA | Karen (Boliviana) | 13,3 | 10,4 | 0,78 | 0,01 | 38 | 0,5 | 0 | -21,22116981 | -44,97728181 | 38 | 4.215 | 6351 | 2 | 0,2 | 0,4% |
| Oceano atlantico norte, ponto zero | Stellarium | | | | | 0 | | 4.210 | 16,52 | -45,62 | 0 | 0 | | | | |

Fonte: dados coletados e processados pelos autores.

A diferença percentual entre os resultados alcançados pelos participantes e o valor teórico do raio da Terra variou bastante no experimento. Tivemos um participante que chegou bem próximo, com 0,1% de diferença apenas, e outra que registrou uma diferença de 42,8%. Essas variações eram esperadas, considerando que muitos fatores podem interferir nas medições. O tempo exato da sombra mínima pode não ter sido captado pelo participante; a sombra pode ter se deformado à medida que o tempo foi passando e o participante pode não ter reposicionado o objeto; há ainda fatores relacionados ao cálculo da distância entre o local do experimento e o ponto zero, entre outros. As variações de resultados enriquecem a experiência e permitem aos participantes compreenderem que o método científico exige uma precisão de procedimentos e constante checagem e repetição nas coletas.

Ao observarmos o perfil dos participantes que concluíram o experimento, verificamos que a maioria foi de estudantes e professores, com apenas dois deles identificando-se com outras profissões. Apesar de o chamado à participação ter se estendido para um público amplo, com repercussão inclusive na imprensa, aqueles que assumiram a tarefa até o fim foram majoritariamente os envolvidos no momento com o ensino formal. Isso sugere que as experiências práticas são bem-vindas para o ambiente de ensino-aprendizagem. No formulário, ao responder à questão sobre o motivo de terem se interessado em realizar o experimento, apareceram respostas que mostram haver um interesse dos participantes pela ciência e pela possibilidade de observação.

*"Sou professora e desejo replicar o experimento com meus alunos e na rede que trabalho".*

*"Porque sou apaixonado pelas ciências naturais, e para estimular minha filha pela ciência".*

*"É uma oportunidade incrível de entender melhor esse experimento que foi realizado séculos atrás e que apesar dos recursos serem escasso naquela época Eratóstenes conseguiu chegar com êxito no resultado final."*

Na questão que buscou apurar a percepção final dos participantes sobre a curvatura do planeta, após o experimento, a maioria das respostas demonstram convicção sobre o fato de a Terra ser esférica, tendo o experimento contribuído para essa percepção. "Sempre acreditei que a Terra tem um formato geóide mas não entendia por completo como tinham chegado a essa conclusão e com esse experimento ficou mais claro", escreveu um dos participantes.

Em apenas duas das respostas houve o indicativo de que o experimento não foi suficiente para a demonstração. Em uma delas, consta: "El proceso nos muestra una variación constante de la sombra y de la posición del Sol, que es muy particular. Nos indica que el proceso tiene inicio y fin para nuestro horizonte local, y que sólo eso podemos registrar y describir. Esta limitación nos impide tener una noción más clara de la curvatura de la Tierra. Necesitamos más información para concluir que la Tierra tiene curvatura y poder cuantificarla".

Essa última percepção, apesar de não adotar o experimento como decisivo para a percepção da Terra esférica, mostram-se abertas ao diálogo e ao aprofundamento das discussões, indicando aos autores a necessidade de realização de novo encontro virtual entre os participantes, para discussão dos resultados. Dessa forma, o experimento foi importante para o processo de interação com as pessoas acerca de temas atacados por proposições negacionistas.

**Conclusões**

Toda a dinâmica do experimento coletivo possibilitou que os participantes e as participantes refletissem sobre a importância do método científico. Até mesmo para aqueles que não chegaram a realizar as medições, acreditamos ter sido válido no sentido de terem podido perceber o quanto a investigação científica é criteriosa e detalhada.

Quando se fala no comprimento da sombra no experimento de Eratóstenes, parece uma instrução muito simples, mas, na experiência, essa sombra não é qualquer sombra - é a mínima sombra. Para conseguir esse dado, o público relatou ter "batalhado com a sombra", porque a cada momento ela se deforma devido ao movimento aparente do sol, o que requer bastante atenção por parte de quem está realizando as medições, conforme disseram.

Este experimento, além de conectar o cotidiano e sua relação com a curvatura da Terra, acabou integrando pessoas de diferentes países, causando uma vivência real nos investigadores e nas investigadoras. Mais impressões sobre a vivência de este experimento comunitário podem ser encontradas no site do projeto [1]. Dessa forma, avaliamos que empreendimentos como este podem ser um recurso importante para o enfrentamento ao negacionismo da ciência, auxiliando o público não especializado em suas reflexões e percepções, especialmente sobre como se dá a aplicação do método científico.

**Referências**